\documentclass[preprint,authoryear,12pt]{elsarticle}

\usepackage{natbib}
\usepackage{graphics,epsfig}
\usepackage{graphicx,epstopdf}
\usepackage{amssymb,amstext,amsmath}

\journal{International Journal of Engineering Science}

\begin{document}

\begin{frontmatter}

\title{Thermal softening during high-temperature torsional deformation of aluminum bars}

\author{K.C. Le$^{a,b}$\footnote{Corresponding author: email: lekhanhchau@tdtu.edu.vn}, Y. Piao$^c$} 

\address{$^a$\,Materials Mechanics Research Group, Ton Duc Thang University, Ho Chi Minh City, Vietnam\\
$^b$\,Faculty of Civil Engineering, Ton Duc Thang University, Ho Chi Minh City, Vietnam\\
$^c$\,Lehrstuhl f\"ur Mechanik - Materialtheorie, Ruhr-Universit\"at Bochum, D-44780 Bochum, Germany}

\begin{abstract} A simple extension of the thermodynamic dislocation theory to non-uniform plastic deformations is proposed for an analysis of high-temperature torsion of aluminum bars. Employing a small set of physics-based parameters, which we expect to be approximately independent of strain rate and temperature, we are able to fit experimental torque-twist curves for five different twist rates and at one fixed ambient temperature. We find that thermal softening due to temperature rise is significant at high twist rates. 
\end{abstract}


\end{frontmatter}

\section{Introduction}

The aim of this paper is to explore use of the thermodynamic dislocation theory  \citep{langer2010thermodynamic,langer2015statistical,langer2016thermal,le2017thermodynamic,le2018thermodynamic,le2018athermodynamic,le2018bthermodynamic,le2018cthermodynamic} in modelling high-temperature torsional deformation of aluminum bars.  We analyze a set of  torsion tests for an aluminum alloy 5252 reported by \citet{zhou1998finite}.  By making torque-twist measurements over a range of substantially different twist rates and at a fixed ambient temperature, these authors have experimentally discovered the thermal softening effect during torsional deformation of aluminum bars, especially at high twist rates.  Our challenge is to predict this effect by using a realistic physics-based theory, and in addition, to obtain basic information about this material.  

Our new ability to interpret data of the kind published in \citep{zhou1998finite} is due to the fact that the thermodynamic dislocation theory, in its latest versions \citep{langer2016thermal,le2017thermodynamic,le2018athermodynamic,le2018bthermodynamic}, includes the non-uniform plastic deformations and the thermal softening. Earlier versions of the theory \citep{langer2010thermodynamic,langer2015statistical} were developed exclusively for uniform plastic deformations. Besides, based on data for highly conductive copper as shown in \citep{kocks2003physics} or \citep{meyers1995effect}, the thermal softening was completely ignored.  \citet{le2018cthermodynamic} have proposed a theory of twisted bars for both macro and micro sizes which shows that the density of excess dislocations is negligible for twisted bars of macro sizes with small to medium twist angles. For such bars a simple extension of the TDT is possible, which takes into account the spatial variation of state variables and ignores the excess dislocations, and was proposed by \citet{le2018bthermodynamic}. To include the thermal effect, we extend that simple version of TDT in this paper by adding the equation for the ordinary kinetic vibrational temperature at which the temperature rise is caused by the movement of dislocations.  With the present theory and experimental data from \citep{zhou1998finite}, the thermal softening effect during torsional deformations can now be systematically investigated.  

The plan of this short communication is as follows. We start in Sec.~\ref{EOM} with a brief annotated summary of the proposed equations of motion. In Sec.~\ref{NI} we discretize the obtained system of governing equations and develop the numerical method for its solution. The parameter identification based on the large scale least squares analysis is presented in Sec.~\ref{PI}. Sec.~\ref{NS} shows the results of numerical simulations and the comparison with experiments. Finally, Sec.~\ref{CS} concludes the paper.

\section{Equations of motion}
\label{EOM}
 
Suppose a single crystal bar with a circular cross section, of radius $R$ and length $L$, is subjected to torsion (see the bar with its cross-section in Fig.~1). For this particular geometry of the bar it is natural to assume that the circumferential displacement is $u_\varphi=\omega rz$, with $\omega $ being the twist angle per unit length, while the displacement $u_z$ does not depend on $\varphi$. Thus, the total shear strain of the bar $\gamma =2\epsilon_{\varphi z}=\omega r$ and the shear strain rate $\dot{\gamma}=\dot{\omega}r$ turn out to be non-uniform as they are  linear functions of radius $r$. Now, let this system be driven at a constant twist rate $\dot\omega \equiv \phi_0/t_0$, where $t_0$ is a characteristic microscopic time scale. Since the system experiences a steady state torsional deformation, we can replace the time $t$ by the total twist angle (per unit length) $\omega$ so that $t_0\,\partial/\partial t \to \phi_0\,\partial/\partial \omega$.

\begin{figure}[t]
	\centering
	\includegraphics[width=.5\textwidth]{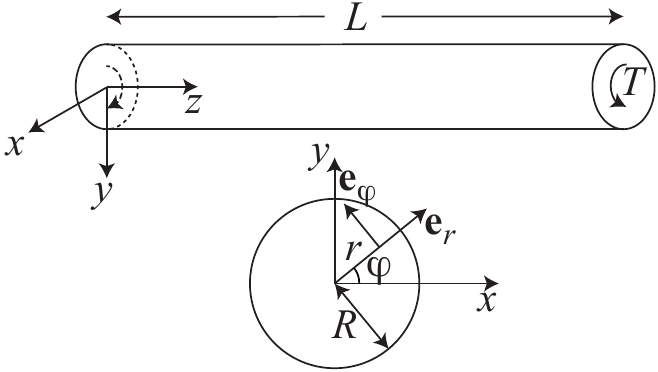}
	\caption{Torsion of a bar}
	\label{bar}
\end{figure}

To derive the equations of motion for this system we begin with Hooke's law in rate form
\begin{equation}
\label{Hooke}
\frac{\partial \tau}{\partial t}=\mu \Bigl( \frac{\partial \gamma}{\partial t} - \frac{\partial \beta }{\partial t}\Bigr) ,
\end{equation}
with $\tau (r,t)=\sigma_{\varphi z}$ being the shear stress, $\beta (r,t)=2\varepsilon^p_{\varphi z}$ the plastic distortion, and $\mu$ the shear modulus. Note that the shear modulus $\mu$ can depend on the ordinary temperature and thus on $t$, yet the term containing $\dot{\mu}$ on the right side of Eq.~\eqref{Hooke} is neglected as small. The central, dislocation-specific ingredient of TDT \citep{langer2010thermodynamic} is the thermally activated depinning formula for the plastic distortion rate, $\dot{\beta}$, as a function of the stress $\tau$ and the total dislocation density $\rho$:  
\begin{equation}
\label{qdef}
\dot{\beta }=\frac{q(\tau,\rho)}{t_0}=\frac{1}{t_0} b\sqrt{\rho} [f_P(\tau,\rho)-f_P(-\tau,\rho)], 
\end{equation}
where
\begin{equation}
\label{exp}
f_P(\tau,\rho)=\exp\,\Bigl[-\,\frac{1}{\theta}\,e^{-\tau/\tau_T(\rho)}\Bigr]. \notag
\end{equation}
As argued in \citep{langer2010thermodynamic}, Eq.~(\ref{qdef}) is an Orowan relation of the form $\dot{\beta} = \rho\,b\,v$ in which the speed of the dislocations $v$ is given by the distance between them multiplied by the rate at which they are depinned from each other. That rate is approximated here by the activation terms $f_P(\tau,\rho)$ and $-f_P(-\tau,\rho)$, in which the energy barrier $e_P=k_BT_P$ (implicit in the scaling of $\theta=T/T_P$) is reduced by the stress dependent factor $e^{-\tau/\tau_T(\rho)}$, where  $\tau_T(\rho)= \mu_T\,b \sqrt{\rho}$ is the Taylor stress with $\mu_T$ being proportional to $\mu(\theta)$ (see Section \ref{NI}). 

Replacing $\dot{\gamma}$ in Eq.~(\ref{Hooke}) by $(\phi_0/t_0)r$, $\dot{\beta}$ by $q(\tau,\rho)/t_0$, and the partial time derivative $\partial/\partial t$ by $(\phi_0/t_0)\,\partial/\partial \omega$, we arrive at
\begin{equation}
\label{tauydot}
\frac{\partial \tau}{\partial \omega} = \mu(\theta) \Bigl[r - \frac{q(\tau,\rho)}{\phi_0}\Bigr].
\end{equation}

The equation of motion for the total dislocation density $\rho$ describes energy flow \citep{langer2010thermodynamic}. In terms of the twist angle $\omega$ this equation reads: 
\begin{equation}
\label{rhodot}
\frac{\partial \rho}{\partial \omega} = K_\rho \,\frac{\tau\,q}{a^2\nu(\theta,\rho,\phi_0r)^2\,\mu(\theta)\,\phi_0}\, \Bigl[1 -\frac{\rho}{\rho_{ss}(\chi)} \Bigr],
\end{equation}
with $\rho_{ss}(\chi) =(1/a^2)e^{- e_D/\chi}$ being the steady-state value of $\rho$ at given $\chi$, $e_D$ a characteristic formation energy for dislocations, and $a$ denoting the average spacing between dislocations in the limit of infinite  $\chi$ ($a$ is a length of the order of tens of atomic spacings). The coefficient $K_\rho $ is an energy conversion factor that, according to arguments presented in \citep{langer2010thermodynamic}, should be independent of both strain rate and temperature. The other quantity that appears in the prefactor in Eq.~(\ref{rhodot}) is
\begin{equation}
\label{nudef}
\nu(\theta,\rho,\phi_0r) \equiv \ln\Bigl(\frac{1}{\theta}\Bigr) - \ln\Bigl[\ln\Bigl(\frac{b\sqrt{\rho}}{\phi_0r}\Bigr)\Bigr].
\end{equation}

The equation of motion for the effective disorder temperature $\chi$ is a statement of the first law of thermodynamics for the configurational subsystem \citep{langer2010thermodynamic}: 
\begin{equation}
\label{chidot}
\frac{\partial \chi }{\partial \omega} = K_\chi \,\frac{\tau e_D \,q}{\mu(\theta)\,\phi_0}\,\Bigl( 1 -\frac{\chi}{\chi_0} \Bigr). 
\end{equation}
Here, $\chi_0$ is the steady-state value of $\chi$ for strain rates appreciably smaller than inverse atomic relaxation times, i.e. much smaller than $t_0^{-1}$. The dimensionless factor $K_\chi $ is inversely proportional to the effective specific heat $c_{e\!f\!f}$. Since the maximum strain rate (reached at the outer radius of the bar) for the small twist rate in our torsion test is small, we assume that $K_\chi$ is a constant. 

The equation of motion for the ordinary temperature $\theta=T/T_P$ reads 
\begin{equation}
\label{thetadot}
\frac{\partial \theta }{\partial \omega} = K_1(\theta)\,\frac{\tau q}{\mu(\theta) \phi_0} - \frac{K_2}{\phi_0}\,(\theta - \theta_0).
\end{equation} 
Here, $K_1(\theta) = \beta \mu(\theta)/ (T_P\,c_p\,\rho_d)$ is a thermal energy conversion factor, with $c_p$ being the thermal heat capacity per unit mass, $\rho_d$ the mass density, and $0< \beta < 1$ a dimensionless constant known as the Taylor-Quinney factor. Thus, the first term on the right-hand side of Eq.~\eqref{thetadot} represents the portion of plastic power dissipated into heat. As indicated here, $K_1(\theta)$ will be found to be temperature dependent. $K_2$ is a thermal transport coefficient that controls how rapidly the system relaxes toward the ambient temperature $T_0$, that is, $\theta \to \theta_0 = T_0/T_P$.  This coefficient turns out to be small for the situations reported here. Nevertheless it cannot be neglected, especially for small twist rates. In principle, after long enough times of steady deformation, systems must reach steady-state temperatures determined by the balance between heating and cooling terms in Eq.~(\ref{thetadot}).   

\section{Discretization and method of solution}
\label{NI}

For the purpose of numerical integration of the system of equations (\ref{tauydot})-(\ref{thetadot}) let us introduce the following dimensionless variables and quantities
\begin{eqnarray}
\tilde{r}=r/R,\quad \tilde{\rho}=a^2\rho ,\quad \tilde{\chi }=\frac{\chi }{e_D},  
\quad \tilde{\omega}=R\omega .\label{dimless}
\end{eqnarray}
The dimensionless variable $\tilde{r}$ changes from zero to $1$. The  dimensionless variable $\tilde{\omega }$ has the meaning of the maximum shear strain achieved at the outer radius. The calculation of the rescaled torque $\tilde{T}=T/R^3$ as function of $\tilde{\omega}=\omega R$ is convenient for the later comparison with the experimental data from \citep{zhou1998finite}. Then we rewrite Eq.~(\ref{qdef}) in the form
\begin{equation}
\label{tildeq}
q(\tau,\rho)=\frac{b}{a}\tilde{q}(\tau,\tilde{\rho}),
\end{equation}
where
\begin{equation}
\label{tildeqdef}
\tilde{q}(\tau,\tilde{\rho})=\sqrt{\tilde{\rho}}[\tilde{f}_P(\tau,\tilde{\rho})-\tilde{f}_P(-\tau,\tilde{\rho})].
\end{equation}
We set $\tilde{\mu}_T=(b/a)\mu_T=\mu(\theta) s$ and assume that $s$ is independent of temperature and strain rate. Then
\begin{equation}
\label{tildefp}
\tilde{f}_P(\tau,\tilde{\rho})=\exp\,\Bigl[-\,\frac{1}{\theta}\,e^{-\tau/(\mu s\sqrt{\tilde{\rho }})}\Bigr].
\end{equation}
We define $\tilde{\phi}_0=(a/b)R\phi_0$ so that $q/(R\phi_0)=\tilde{q}/\tilde{\phi}_0$. Function $\nu $ in Eq.~(\ref{nudef}) becomes
\begin{equation}
\label{nudef1}
\tilde{\nu}(\theta,\tilde{\rho},\tilde{\phi}_0\tilde{r}) \equiv \ln\Bigl(\frac{1}{\theta}\Bigr) - \ln\Bigl[\ln\Bigl(\frac{\sqrt{\tilde{\rho}}}{\tilde{\phi}_0\tilde{r}}\Bigr)\Bigr].
\end{equation}
The dimensionless steady-state quantities are
\begin{equation}
\label{ss}
\tilde{\rho}_{ss}(\tilde{\chi})=e^{-1/\tilde{\chi}}, \quad \tilde{\chi}_0=\chi_0/e_D.
\end{equation}
Using $\tilde{q}$ instead of $q$ as the dimensionless measure of plastic strain rate means that we are effectively rescaling $t_0$ by a factor $b/a$. Since $t_0^{-1}$ is a microscopic attempt frequency, of the order $10^{12}$\,s$^{-1}$, we take $(a/b)t_0=10^{-12}$s.

In terms of the introduced quantities the governing equations read
\begin{eqnarray}
\frac{\partial \tau}{\partial \tilde{\omega}} = \mu(\theta) \Bigl[ \tilde{r} - \frac{\tilde{q}(\tau,\tilde{\rho})}{\tilde{\phi }_0}\Bigr], \label{tau} 
\\
\frac{\partial \tilde{\rho}}{\partial \tilde{\omega}} = K_\rho \,\frac{\tau\,\tilde{q}}{\mu(\theta) \tilde{\nu}(\theta,\tilde{\rho},\tilde{\phi}_0\tilde{r})^2\,\tilde{\phi}_0}\, \Bigl[1 -\frac{\tilde{\rho}}{\tilde{\rho}_{ss}(\tilde{\chi})} \Bigr], \label{rho}
\\
\frac{\partial \tilde{\chi }}{\partial \tilde{\omega}} = K_\chi\,\frac{\tau\,\tilde{q}}{\mu(\theta) \tilde{\phi }_0}\,\Bigl( 1 -\frac{\tilde{\chi}}{\tilde{\chi}_0} \Bigr) , \label{chi}
\\
\frac{\partial \theta}{\partial \tilde{\omega}} = K_1(\theta)\,\frac{\tau\,\tilde{q}}{\mu(\theta) \tilde{\phi }_0}-\frac{\tilde{K}_2}{\tilde{\phi}_0}\, (\theta -\theta_0) , \label{theta}
\end{eqnarray}
where $\tilde{K}_2=(a/b)K_2$. To solve this system of differential equations subject to initial conditions numerically, we discretize the equations in the interval $(0< \tilde r <1)$ by dividing it into $n$ sub-intervals of equal length $\Delta \tilde{r}=1/n$ and writing the corresponding equations at $n$ nodes $\tilde{r}_i=i \Delta \tilde{r}$, $i=1,\ldots,n$. In this way,  we reduce the four differential equations depending on $\tilde{r}$ to a system of $4n$ ordinary differential equations at $n$ nodes that will be solved by Matlab-ode23s. 
 
After finding the solution we can compute the torque as function of the twist angle according to
\begin{equation}
\label{torque}
T = 2\pi R^3 \int_0^{1} \tau \tilde{r}^2 d\tilde{r}.
\end{equation}

\section{Parameter identification}
\label{PI}

In order to simulate the theoretical torque-twist curves, we need values for eight system-specific parameters and two initial conditions from each sample.  These basic parameters include the following six: the activation temperature $T_P$, the stress ratio $s$, the steady-state scaled effective temperature $\tilde\chi_0$, and the three dimensionless conversion factors $K_\rho$, $K_\chi$, and $\tilde{K}_2$. In addition, we need a formula for the thermal conversion factor $K_1(\theta)$ in Eq.~(\ref{thetadot}) which is assumed to be a linear function of $\theta$  
\begin{equation}
\label{Ktheta}
K_1(\theta) = K_0 \,\left[1 + c_1\,T_P\,(\theta - \theta_0)\right].
\end{equation}
The numbers $K_0$ and $c_1$ are the two remaining parameters to be determined from the data. We also need initial values of the scaled dislocation density $\tilde\rho_i$ and the effective disorder temperature $\tilde\chi_i$; all of which are determined by the sample preparation. Concerning the initial values of the shear stress and the ordinary temperature we assume that $\tau_i=0$ and $\theta_i=\theta_0$, where $T_0=T_P\, \theta_0=773$\,K is the ambient temperature. The other parameters required for numerical simulations but known from the experiment are: the length $L=10$\,mm and radius $R=5$\,mm of the bars, and the length of Burgers' vector $b=2.86$\,\AA. Since $a$ corresponds to the smallest admissible distance between dislocations in the state of maximum disorder in crystal, we take $a=10b$. Note that $a$ only affects the dislocation density, not the torque-twist curves. Finally, for the temperature dependent shear modulus $\mu(\theta)$ we use a formula proposed by \citet{varshni1970temperature}  
\begin{equation}
\label{muAl}
\mu(\theta) = \mu_1 - \Bigl[\frac{D}{\exp(T_1/T_P\, \theta)-1}\Bigr],
\end{equation}
where $\mu_1 = 28.8$\,GPa, $D = 3.44$\,GPa, and $T_1 = 215$\,K. (These values taken from \citep{chen1998constitutive} is actually for the aluminum alloy 5182, but since the chemical composition of the aluminum alloy 5252 is very close to that of 5182, we use the same parameters for our alloy.)

In earlier papers dealing with the uniform deformations \citep{langer2010thermodynamic,langer2015statistical}, it was possible to begin evaluating the parameters by observing steady-state stresses $\tau_{ss}$ at just a few strain rates $q_0$ and ambient temperatures $T_0 = T_P\,\theta_0$. Knowing $\tau_{ss}$, $T_0$ and $q_0$ for three stress-strain curves, one could solve equation 
\begin{equation}
\label{qdef2}
\tau = \tau_T(\tilde\rho)\,\nu(\theta,\tilde\rho,q_0),
\end{equation} 
which is the inverse of Eq.~(\ref{qdef}) for $T_P$, $s$, and $\tilde\chi_0$, and check for consistency by looking at other steady-state situations. With that information, it was relatively easy to evaluate $K_\rho$ and $K_\chi$ by directly fitting the full stress-strain curves.  This strategy does not work here because the stress state of twisted bars is non-uniform. Besides, the thermal effects are highly nontrivial.  Examination of the experimental data provided in \citep{zhou1998finite} indicates that almost all of these samples are undergoing thermal softening at large twist angles; the torques are decreasing and the temperatures must be increasing.  Even the curves that appear to have reached some kind of steady state have not, in fact, done so at their nominal ambient temperatures. 
 
\begin{table}
  \centering 
\begin{tabular}{|c|c|c|c|c|c|}
\hline
$R\dot{\omega}$\,(1/s) & 19.7 & 6.44 & 1.84 & 0.21 & 0.02 \\ \hline
$\tilde{\rho}_i$ & $7.94\times 10^{-5}$ & $9.49\times 10^{-5}$ & $7.66\times 10^{-5}$ & $7.59\times 10^{-5}$ &  $2.03\times 10^{-5}$\\  
\hline
$\tilde{\chi}_i$ & 0.245 & 0.24 & 0.234 & 0.23 & 0.2235 \\ \hline
\end{tabular}  
  \caption{The initial values of $\tilde{\rho}_i$ and $\tilde{\chi}_i$}\label{table:1}
\end{table}

To counter these difficulties, we have resorted to the large-scale least-squares analyses that we have developed in \citep{le2017thermodynamic,le2017dislocation,le2018bthermodynamic,le2018cthermodynamic}. That is, we have solved the system of ordinary differential equations (ODEs) numerically, provided a set of material parameters and initial values is known. Based on this numerical solutions we then computed the sum of the squares of the differences between our theoretical torque-twist curves and a large set of selected experimental points, and minimized this sum in the space of the parameters. The ODEs were solved numerically using the Matlab-ode23s, while the finding of least squares was realized with the Matlab-globalsearch. To keep the calculation time manageable and simultaneously ensure the accuracy, we have chosen $n=100$ and the $\tilde{\omega}$-step equal to $\tilde{\omega}_*/8000$. Our results appear to be robust against the experimental uncertainties. We have found that the torque-twist curves for five samples taken from \citep{zhou1998finite} can be fit with just a single set of system parameters. These are: $T_P = 23628$\,K,\,$s = 0.052,\,\chi_0= 0.2489,\,K_\rho=10.16,\, K_\chi=327.72,\, K_0=0.152,\, c_1=0.0057,\, \tilde{K}_2=1.762\times 10^{-13}$. So far as we can tell, our values of $K_0$ and $c_1$ are consistent with values of the Taylor-Quinney factor $\beta$ of the order of unity or less, at least in the range $T<900$\,K (melting temperature). The identified initial values of dislocation densities and disorder temperatures for five samples are shown in Table 1, where the first row indicates the twist rates at which the samples are twisted. To obtain the actual initial dislocation densities, we must divide $\tilde{\rho}_i$ by $a^2$, resulting in the order between $10^{12}$ and $10^{13}$ dislocations per square meter.  

\section{Numerical simulations}
\label{NS}

\begin{figure}[htp]
	\centering
	\includegraphics[width=.65\textwidth]{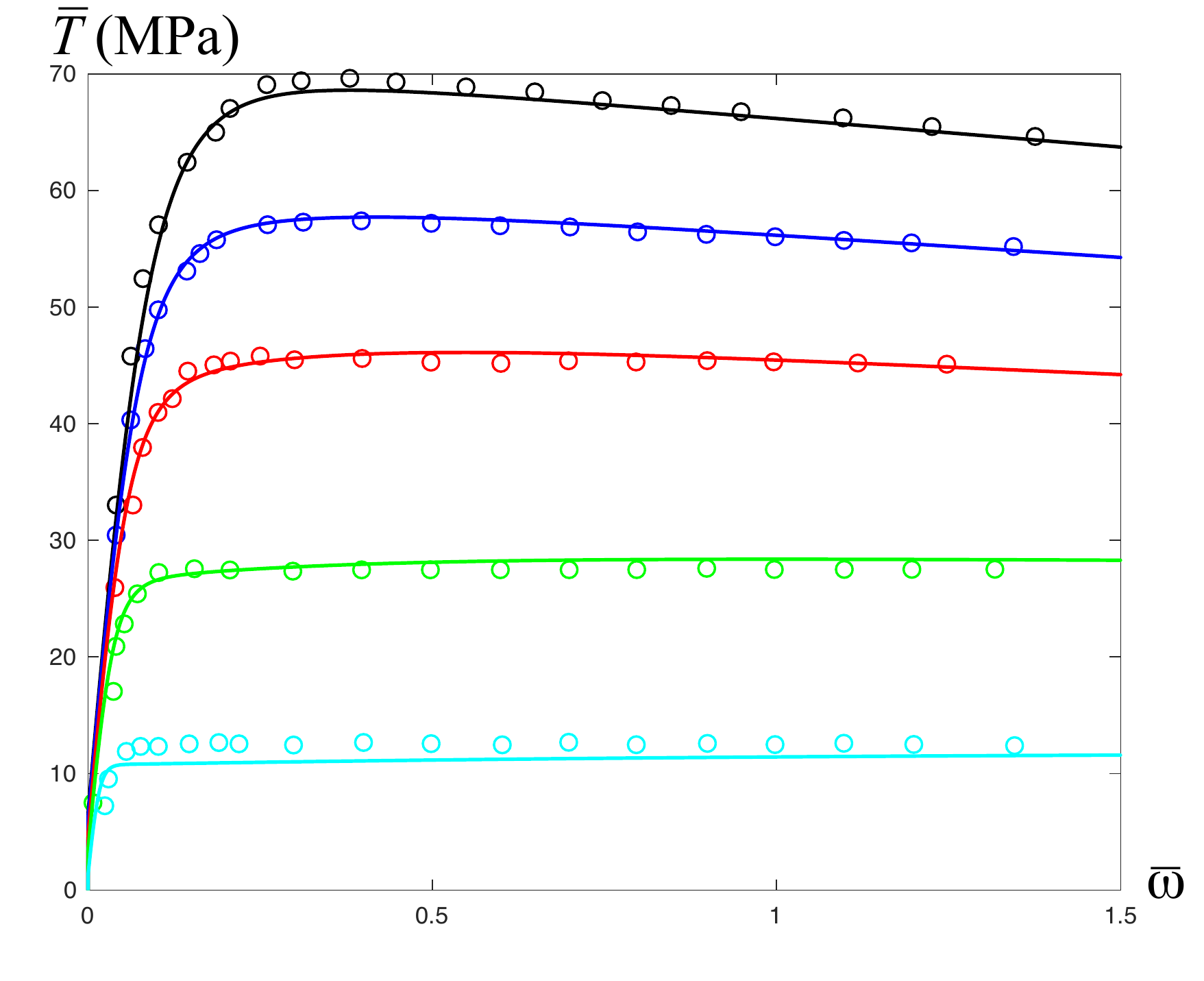}
	\caption{(Color online) Torque-twist curves for ambient temperature $T_0=773$\,K, and for five different twist rates $R\dot{\omega}=19.7/$s (black), $R\dot{\omega}=6.44/$s (blue), $R\dot{\omega}=1.84/$s (red), $R\dot{\omega}=0.21/$s (green) and $R\dot{\omega}=0.02/$s (cyan): (i) TDT (bold lines), (ii) experimental points taken from \citet{zhou1998finite} (circles).}
	\label{TorqueTwist}
\end{figure}

With the identified parameters we can now simulate the torque-twist curves for bars made of aluminum alloy 5252 undergoing torsional deformations at five different twist rates $R\dot{\omega}=19.7/$s, $6.44/$s, $1.84/$s, $0.21/$s, and $0.02/$s, and for ambient temperature $T_0=773$\,K. In oder to compare with the experimental curves we use the rescaled torque and twist angle (or effective strain) defined as followed \citep{zhou1998finite}
\begin{equation}
\label{rescaled}
\bar{T}=\frac{12\sqrt{3}}{8\pi R^3}T, \quad \bar{\omega}=\frac{0.722R\omega}{\sqrt{3}}.
\end{equation}
The result is presented in Fig.~\ref{TorqueTwist}. In this figure, the circles represent the selected experimental points in \citep{zhou1998finite} while the solid curves are our theoretical simulation. One can see that even the initial yielding transition appears to be described accurately by this theory. There is only one visible discrepancy: for the smallest twist rate $R\dot{\omega}=0.02/$s the torques are slightly above those predicted by the theory. Nothing about this result leads us to believe that there are relevant physical ingredients missing in the theory. 

\begin{figure}[htp]
	\centering
	\includegraphics[width=.65\textwidth]{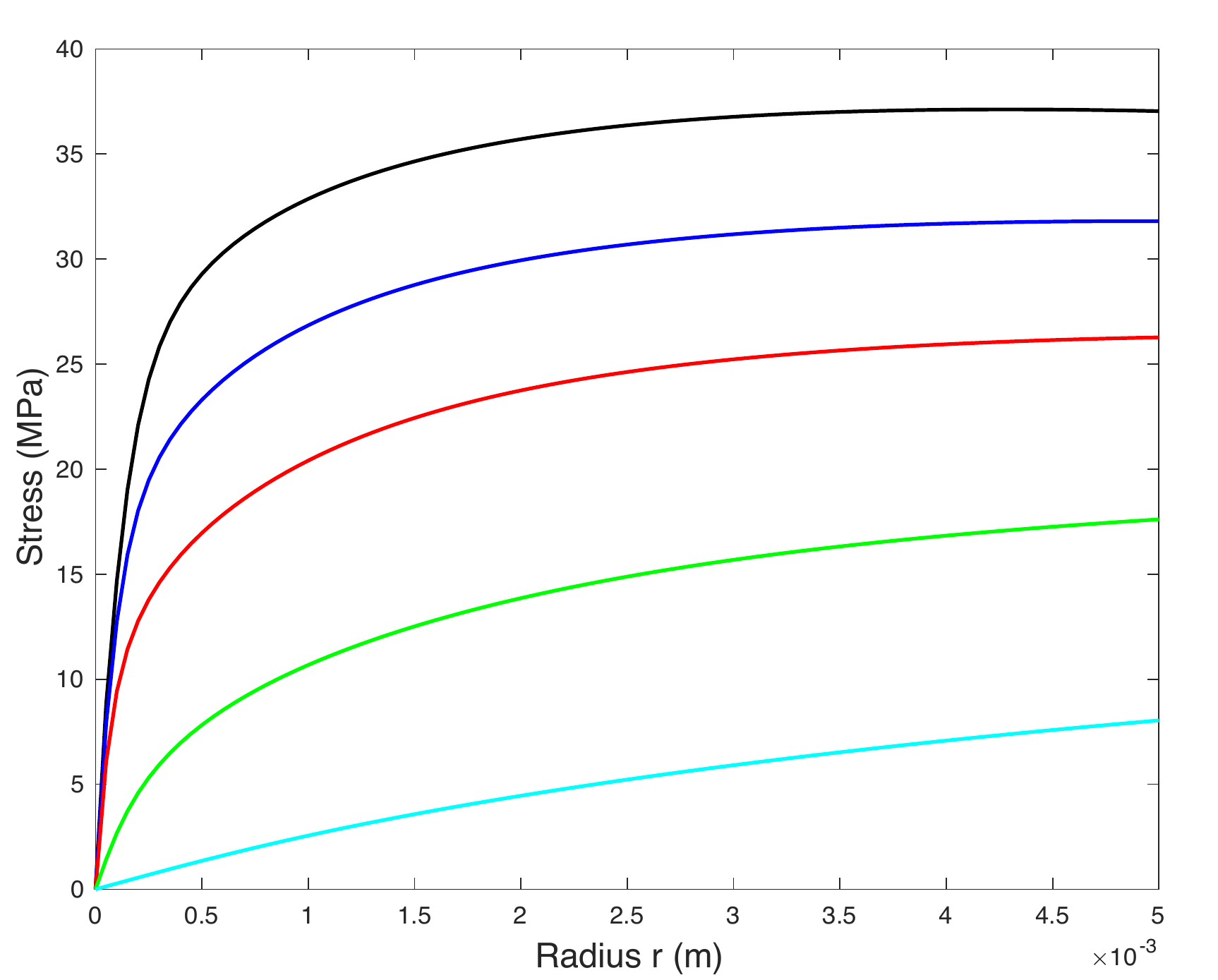}
	\caption{(Color online) Stress distributions $\tau(r)$ at $\bar{\omega}_*=1.5$ for ambient temperature $T_0=773$\,K, and for five different twist rates $R\dot{\omega}=19.7/$s (black), $R\dot{\omega}=6.44/$s (blue), $R\dot{\omega}=1.84/$s (red), $R\dot{\omega}=0.21/$s (green) and $R\dot{\omega}=0.02/$s (cyan).}
	\label{Stress}
\end{figure}

\begin{figure}[htp]
	\centering
	\includegraphics[width=.65\textwidth]{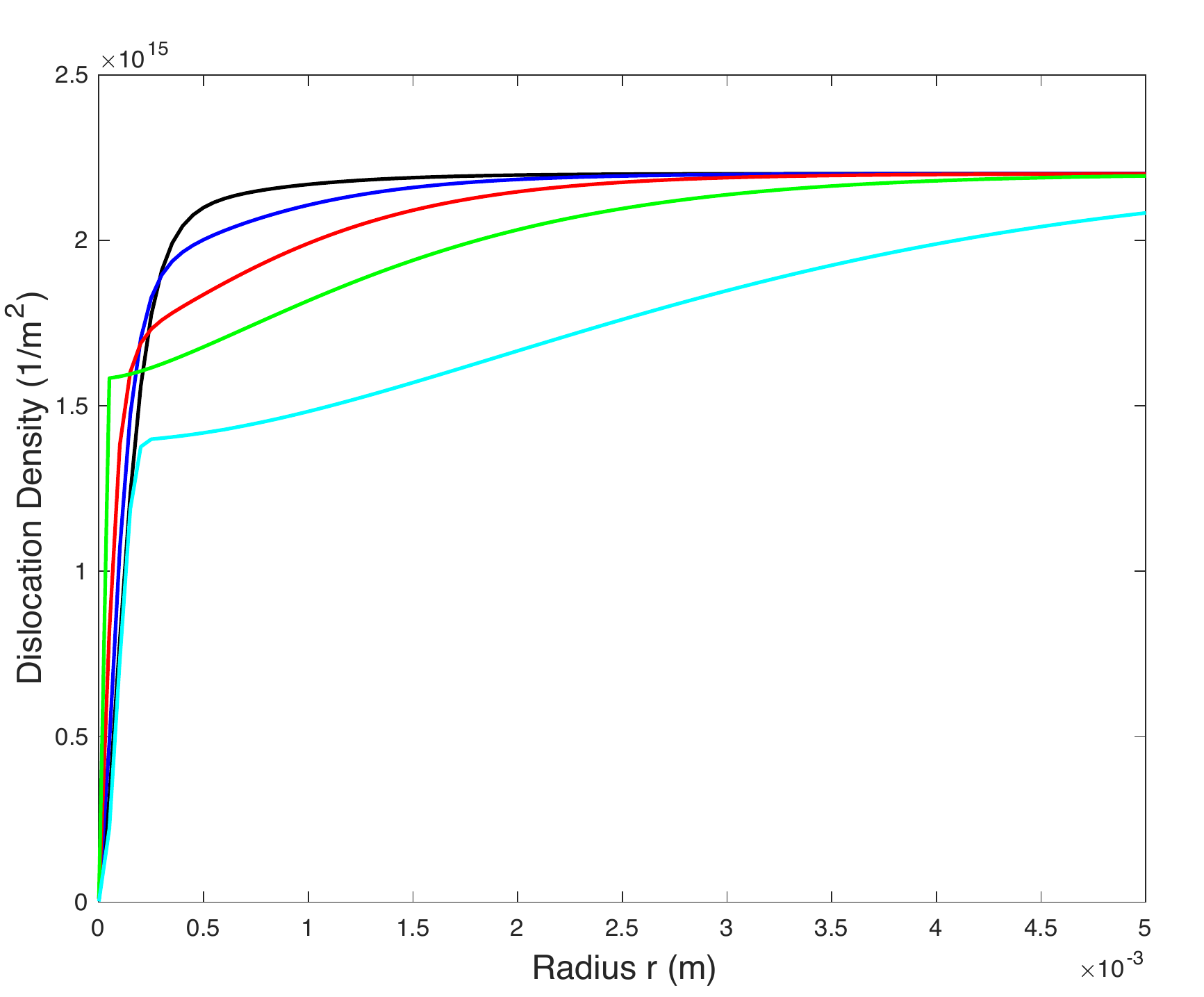}
	\caption{(Color online) Density of dislocations $\rho(r)$ at $\bar{\omega}_*=1.5$ for ambient temperature $T_0=773$\,K, and for five different twist rates $R\dot{\omega}=19.7/$s (black), $R\dot{\omega}=6.44/$s (blue), $R\dot{\omega}=1.84/$s (red), $R\dot{\omega}=0.21/$s (green) and $R\dot{\omega}=0.02/$s (cyan).}
	\label{Densitytotal}
\end{figure}

\begin{figure}[htp]
	\centering
	\includegraphics[width=.65\textwidth]{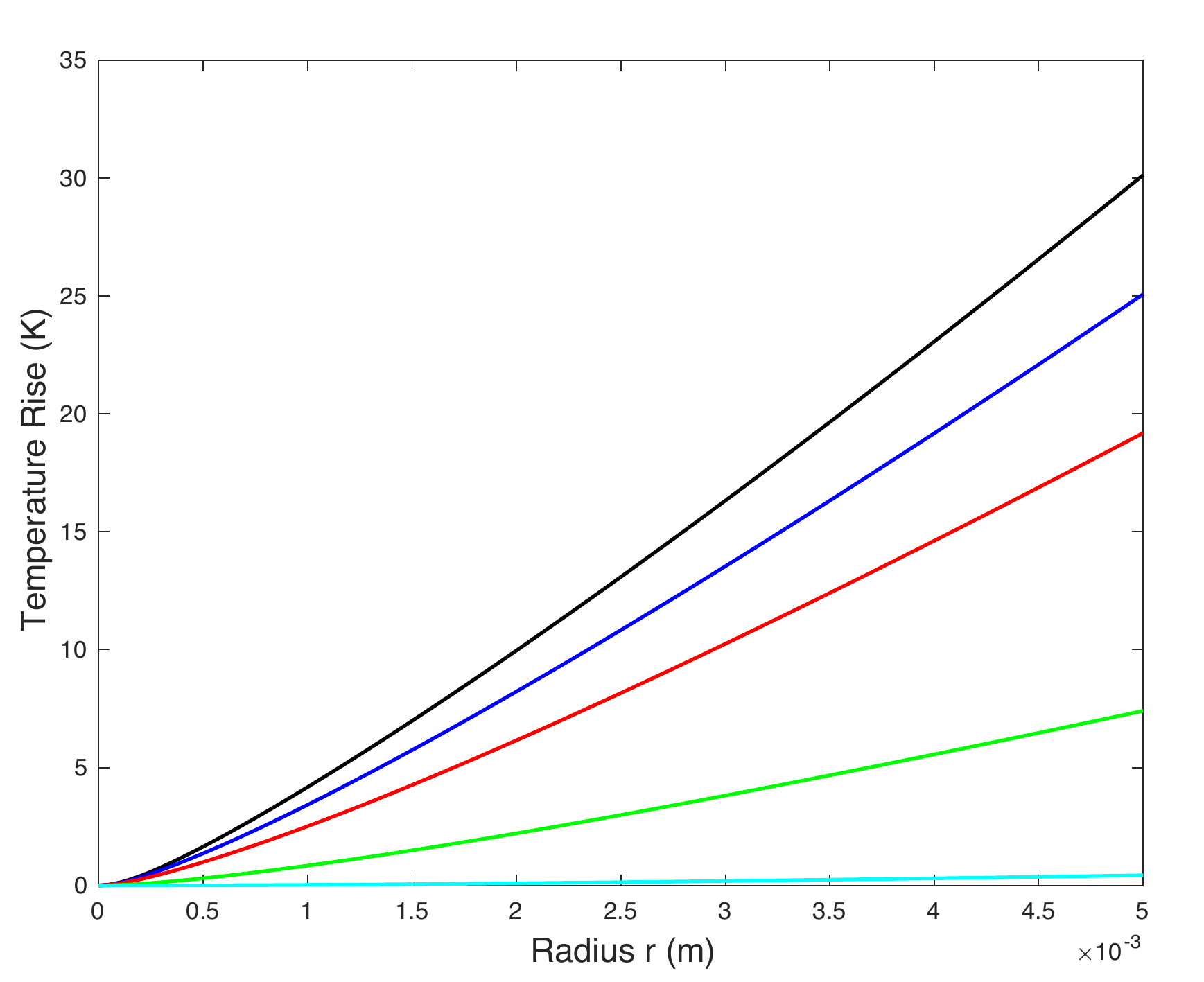}
	\caption{(Color online) Temperature rise $\Delta T(r)$ at $\bar{\omega}_*=1.5$ for ambient temperature $T_0=773$\,K, and for five different twist rates $R\dot{\omega}=19.7/$s (black), $R\dot{\omega}=6.44/$s (blue), $R\dot{\omega}=1.84/$s (red), $R\dot{\omega}=0.21/$s (green) and $R\dot{\omega}=0.02/$s (cyan).}
	\label{TemperatureRise}
\end{figure}

The results of numerical simulations for other quantities are shown in Figs.~\ref{Stress}-\ref{TemperatureRise}. We plot in Fig.~\ref{Stress} the shear stress distribution $\tau(r)$ at the maximal twist angle $\bar{\omega}_*=1.5$. In a small elastic zone near the center of the cross-section, the stress depends linearly on $r$. In the plastic zone, the stress does not remain constant, but increases with increasing $r$ and reaches a maximum at $r=R$, as opposed to the similar distribution obtained by the phenomenological theory of ideal plasticity.  This exhibits the hardening behavior due to the entanglement of dislocations. Fig.~\ref{Densitytotal} presents the distributions of density of dislocations $\rho(r)$ at the maximal twist angle $\bar{\omega}_*=1.5$. The density of dislocations is an increasing function of $r$ but quickly achieves a nearly constant value in the outer ring $r_0<r<R$, especially for the high twist rates.

Finally, we present in Fig.~\ref{TemperatureRise} the distribution of the temperature rise $\Delta T(r)$ at the maximal twist angle $\bar{\omega}_*=1.5$. The temperature rise is a monotonously increasing function of the radius and achieves the maximum at $r=R$. We see that the higher the twist rate, the higher is the temperature rise. At the highest twist rate $R\dot{\omega}=19.7/$s the temperature rise at the outer radius is about 30\,K. Our theory predicts that the larger temperature increases occur at the higher twist rates because the plastic power is larger there. This is confirmed by the calculations of the temperature distribution based on the finite element method provided in \citep{zhou1998finite}.

\section{Conclusion}
\label{CS}
Overall, these results seem to us to be quite satisfactory.  Note that we are now using thermodynamic dislocation theory not only to verify its validity, but also as a tool to discover the properties of structural materials.  Thus, at the beginning of this investigation, we did not know that thermal softening would play such an important role for the bars subjected to torsional deformations. One of the main reasons for the success of this theory - as emphasized here and in earlier papers - is the extreme sensitivity of the plastic strain rate to small temperature or stress changes. This allows the yielding transition, work hardening, and thermal softening to be correctly captured, as can be seen in the torque-twist curves. The proposed theory can serve as a useful guide for future experimental investigation of the sensitivity of torque-twist curves to twist rate and ambient temperature. In particular, it would be very interesting to perform torsion tests with extremely high twist rates when the samples are bonded to Kolsky bars (cf. \citep{marchand1988experimental}). In this case, our theory would predict the runaway instability and shear band formation due to the thermal softening. 

\bigskip
\noindent {\it Acknowledgments}

\smallskip
Y. Piao acknowledges financial support from the Chinese Government Scholarship Program. K.C. Le is grateful to J.S. Langer for helpful discussions.

\end{document}